\documentclass[letterpaper,english,dvips]{emulateapj}

\usepackage{graphicx}
\usepackage{amssymb}
\usepackage{natbib}
\usepackage{amsmath}
\usepackage{color}

\makeatletter

\slugcomment{}
\shorttitle{}
\shortauthors{}

\providecommand{\tabularnewline}{\\}

\newcommand{\Kelvin}{\mathrm{K}}
\newcommand{\Msun}{\mathrm{M_{\sun}}}

\makeatother

\begin{document}

\def\MSUN{\rm M_{\odot}}
\def\RSUN{\rm R_{\odot}} 
\def\MSUNYR{\rm M_{\odot}\,yr^{-1}}
\def\MDOT{\dot{M}}

\newbox\grsign \setbox\grsign=\hbox{$>$} \newdimen\grdimen \grdimen=\ht\grsign
\newbox\simlessbox \newbox\simgreatbox
\setbox\simgreatbox=\hbox{\raise.5ex\hbox{$>$}\llap
     {\lower.5ex\hbox{$\sim$}}}\ht1=\grdimen\dp1=0pt
\setbox\simlessbox=\hbox{\raise.5ex\hbox{$<$}\llap
     {\lower.5ex\hbox{$\sim$}}}\ht2=\grdimen\dp2=0pt
\def\simgreat{\mathrel{\copy\simgreatbox}}
\def\simless{\mathrel{\copy\simlessbox}}

\title{On the Feedback Efficiency of Active Galactic Nuclei}

\author{Ryuichi Kurosawa\altaffilmark{1}, Daniel Proga, and Kentaro Nagamine}

\affil{Department of Physics and Astronomy, University of Nevada Las Vegas,
Box~454002, 4505~Maryland Pkwy, Las Vegas, NV 891541-4002}

\altaffiltext{1}{present address: Department of Astronomy, Cornell
  University, Ithaca, NY 14853-6801}

\email{\{rk,dproga,kn\}@physics.unlv.edu}

\begin{abstract}
We measure and analyze 
the energy, momentum, and mass feedback efficiencies 
due to radiation from active galactic nuclei (AGN) 
\textcolor{black}{in relatively large scale outflows (from $\sim0.01$ 
to $\sim 10$\,pc)}.
Our measurements are based on the two-dimensional (axisymmetric) and time-dependent
radiation-hydrodynamical simulations recently presented in Kurosawa \& Proga. 
In that paper, we studied outflows from a slowly rotating (sub-Keplerian) infalling gas driven
by the energy and pressure of the radiation emitted by the AGN.
These simulations follow dynamics of gas  under the influence of the gravity 
of the central $10^8~\MSUN$  black hole on scales from $\sim0.01$ 
to $\sim 10$~pc. They self-consistently couple 
the accretion-luminosity with the mass inflow rate at the smallest radius
(our proxy for the mass-accretion rate, $\dot{M}_{\mathrm{a}}$). Over
thirty simulations have been performed to investigate how
the results depend on the gas density at the outer radius, $\rho_{\mathrm{o}}$.
A key feature of these simulations is that the
radiation field and consequently the gas dynamics are
axisymmetric, but not spherically symmetric. 
Therefore, the gas inflow and outflow
can occur at the same time. We compare our $\dot{M}_{\mathrm{a}}$-$\rho_{\mathrm{o}}$
relation with that predicted by the Bondi accretion model.  
For high luminosities comparable to the Eddington
limit, the power-law fit
($\dot{M}_{\mathrm{a}} \propto \rho_{\mathrm{o}}^{q}$) to our models
yields  $q\approx 0.5$ instead of $q=1.0$ 
which is predicted by the Bondi model.  
This difference is caused by the outflows which are important for
the overall mass budget at high luminosities.  
The maximum momentum and mass feedback efficiencies found in our models 
are $\sim 10^{-2}$ and $\sim 10^{-1}$, respectively.
However, the outflows are much less important energetically:
the thermal and kinetic powers in units of the radiative luminosity
are  $\sim 10^{-5}$ and $\sim 10^{-4}$, respectively. In addition, the efficiencies
do not increase monotonically with the accretion luminosity but
rather peak around the Eddington limit beyond which a
steady state disk-wind-like solution exists.
Our energy feedback efficiencies are significantly lower than 0.05, which is
required in some cosmological and galaxy merger simulations. 
The low feedback efficiencies found here could have significant
implications on the mass growth of super massive black holes in the
early universe. 
\textcolor{black}{We stress however that we have not considered the
  innermost parts of the accretion and outflow where radiation and
  matter interact most strongly. The feedback from this region could
  have efficiencies significantly above the low values found here.
}

\end{abstract}


\section{Introduction}

\label{sec:Introduction}

The central location of AGN in their host galaxies and the fact that they 
can produce a large amount of energy imply that AGN can play a very important 
role in setting the physical conditions in
their vicinity as well as on larger, galactic and even intergalactic scales
(e.g., \citealt{Igumenshchev:1993}; \citealt{ciotti:1997}, \citeyear{ciotti:2001},
\citeyear{ciotti:2007}; \citealt{king:2003}; \citealt{Murray:2005};
\citealt{Sazonov:2005}; \citealt{Springel:2005b};
\citealt{Begelman:2005}; \citealt{Hopkins:2005}; \citealt{Wang:2006};
\citealt{Thacker:2006}; \citealt{Fabian:2006b},
\citeyear{Fabian:2008}; \citealt{Pelupessy:2007};
\citealt{Krolik:2007}; \citealt{Merloni:2008}; \citealt{Booth:2009},
and references therein). There are many  
indications that support this idea. For example, the presence of broad 
and narrow emission lines, broad and narrow absorption lines, 
in AGN spectra suggests that AGN continuum radiation affects 
the immediate environment
of AGN (see \citealt{Krolik:1999} for an overview).
In addition, the tight correlation between the mass ($M_\mathrm{BH}$)
of the central black hole (BH) in a galactic nucleus and 
the velocity dispersion $\sigma$ of the galaxy's bulge or 
spheroid, the so-called ``$M_{\mathrm{BH}}-\sigma$''  relation
(e.g., \citealt{Ferrarese:2000}; \citealt{Gebhardt:2000}; \citealt{Tremaine:2002}) 
can be explained by the feedback between 
AGN and  the infalling material from large distances. This feedback 
can quench both BH accretion and star formation in the galaxy when BH
reaches a certain mass. AGNs could provide such feedback because they
are very powerful sources of  energy and momentum
(e.g., \citealt{Silk:1998}; \citealt{Blandford:1999}; \citealt{Sazonov:2005}; 
\citealt{Fabian:1999}; \citealt{Fabian:2002}; \citealt{king:2003};
\citealt{Scannapieco:2004}; 
\citealt{Murray:2005}; \citealt{Springel:2005b};
\citealt{DiMatteo:2005}; \citealt{Booth:2009}). 

AGN are powered by mass accretion onto a super massive black hole (SMBH).
To illustrate how the growth of SMBH can be self-regulated
and how AGN feedback can be characterized, let us first express
the radiation luminosity due to accretion as  
\begin{equation}
L_\mathrm{a}= \epsilon_{\mathrm{r}} c^2 \MDOT_{\mathrm{a}},
\label{eq:L-acc}
\end{equation}
where we invoke the simplest assumption such that the luminosity is
proportional to the mass accretion rate ($\MDOT_{\mathrm{a}}$) and a
radiative (or the rest-mass conversion) efficiency
($\epsilon_{\mathrm{r}}$).  Both $\epsilon_{\mathrm{r}}$ and
$\MDOT_\mathrm{a}$ are uncertain.  For example,
$\epsilon_{\mathrm{r}}$ ranges from $\sim 10^{-1}$ in a standard,
radiatively efficient thin disk to $\sim 10^{-11}$ for spherically
symmetric accretion from a low density medium (e.g.,
\citealt{shakura:1973}; \citealt{Shapiro:1973};
\citealt{Meszaros:1975}; \citealt{Soltan:1982}; \citealt{Yu:2002}) while the
mass accretion rate depends on poorly constrained physical conditions and
geometry at large distances from the black hole.

A common method to estimate $\MDOT_\mathrm{a}$ is to adopt the analytic 
formula by \citet{Bondi:1952} who considered spherically symmetric accretion from 
a non-rotating polytropic gas with uniform density $\rho_\infty$
and sound speed $c_\infty$ at infinity. Under these assumptions, 
a steady state solution to the equations 
of mass and momentum conservation exists 
with a mass accretion rate of
\begin{equation}
\MDOT_\mathrm{B}= \lambda \, 4 \pi r^2_\mathrm{B} \rho_\infty c_\infty,
\label{eq:mdot_bondi}
\end{equation} 
where $\lambda$ is a dimensionless parameter
that, for the Newtonian potential, depends only on the adiabatic index
(cf.~\citealt{Bondi:1952}; \citealt{Shu:1992}; \citealt{Frank:1992}). 
The Bondi radius, $r_\mathrm{B}$, is defined as
\begin{equation}
r_\mathrm{B}=\frac{G M}{c^2_\infty}
\label{eq:r_bondi}
\end{equation}
where $G$ is the gravitational constant and $M$ is the mass of the accretor.

One can quantify AGN feedback by measuring 
its efficiency in affecting
the flow of energy, momentum, and mass. In this work, 
we consider only the energy and momentum carried out by matter.
The total energy feedback 
efficiency $\epsilon_{\mathrm{t}}$
is defined as the ratio between the accretion luminosity of the system
$L_{\mathrm{a}}$  (Eq.~{[}\ref{eq:L-acc}]) and the sum of the kinetic
power (kinetic energy flux) $P_{\mathrm{k}}$ and thermal energy power (thermal
energy flux) $P_{\mathrm{th}}$, i.e., 
\begin{equation}
  \epsilon_{\mathrm{t}}=\left(P_{\mathrm{k}}+P_{\mathrm{th}}\right)/L_{\mathrm{a}}\,.
  \label{eq:eff-total-energy}
\end{equation}

Similarly, the kinetic energy and thermal
energy feedback efficiencies ($\epsilon_{\mathrm{k}}$ and $\epsilon_{\mathrm{th}}$)
are defined as
\begin{equation}
  \epsilon_{\mathrm{k}}=P_{\mathrm{k}}/L_{\mathrm{a}}
  \label{eq:eff-kinetic-energy}
\end{equation}
 and 
\begin{equation}
  \epsilon_{\mathrm{th}}=P_{\mathrm{th}}/L_{\mathrm{a}}\,,
  \label{eq:eff-thermal-energy}
\end{equation}
respectively. From these definitions, it obviously follows that
$\epsilon_{\mathrm{t}}=\epsilon_{\mathrm{k}}+\epsilon_{\mathrm{th}}$. 

The momentum feedback efficiency ($\epsilon_{\mathrm{p}}$) is defined
as the ratio of the total wind momentum $p_{\mathrm{w}}$ 
to the total radiation momentum ($L_{\mathrm{a}}/c$), i.e.,
\begin{equation}
  \epsilon_{\mathrm{p}}=p_{\mathrm{w}}/\left(L_{\mathrm{a}}/c\right)\,.
  \label{eq:eff-momentum}
\end{equation}
Lastly, the mass feedback efficiency $\epsilon_{\mathrm{m}}$ is
defined as the ratio of the mass-outflow rate at the outer boundary
$\dot{M}_{\mathrm{out}}$ to the mass-inflow
rate at the inner boundary $\dot{M}_{\mathrm{in}}$, i.e.,
\begin{equation}
  \epsilon_{\mathrm{m}}=
  \dot{M}_{\mathrm{out}}/\dot{M}_{\mathrm{in}}\,.
  \label{eq:eff-mass}
\end{equation}

The most advanced studies of feedback effects were carried out by 
\citet{Springel:2005b} (SDH05 hereafter), \citet{DiMatteo:2005} and
\citet{Booth:2009} (BS09 hereafter).
These studies used computer 
simulations of merging galaxies in which they linked local and global 
processes. This was possible because they adopted relatively
crude phenomenological realizations of star formation, radiative cooling
in a complex multi-phase medium, BH accretion and feedback, and
because their spatial resolution is larger than $r_\mathrm{B}$.
In particular, they assumed values of the above introduced efficiencies
instead of directly computing them.
A main result of these simulations is that the $M_{\mathrm{BH}}-\sigma$ relation can
be reproduced remarkably well and that the relation is insensitive
to the gas fraction in the model galaxies. In addition, the BH mass 
appears to be little
affected by the details of star formation and supernova feedback.

\citet{Begelman:2005} argued that the results from simulations 
of merging galaxies
suggest that 
the feedback regulating BH accretion operates on local scales, 
comparable to $r_\mathrm{B}$ or closer in, rather
than solely on the global scales usually considered 
(see also \citealt{Murray:2005}). 
They also presumed that the insensitivity to gas fraction occurs 
in the galaxy merger simulations because the gas mass is somehow ``maximized'' on the 
scales where the accretion rate is determined. Thus 
on these scales, BH feedback can be much more important than that due to
stars. If this is correct then for state-of-the-art models,
the key feedback processes represent ``subgrid'' physics. 
This is a limitation of current models of AGN feedback in large scale
cosmological and galaxy merger simulations because they cannot be
directly related to AGN physics. 

On the other hand, simulations that aim to provide insights to AGN
physics do not include galaxy but rather focus on $r_\mathrm{B}$ or
even smaller scales (e.g., \citealt{Proga:2007b};
\citealt{Proga:2008}; \citealt{Kurosawa:2008};
\citealt{Kurosawa:2009b, Kurosawa:2009}). Thus they cannot be
directly related to AGN feedback on large scales. However, these
smaller scale simulations can be used directly to measure the feedback 
efficiencies listed above.
Consequently, they can be used to quantify the effects that 
are assumed or parametrized in large scale simulations.

The goal of this paper is to present measurements of the mass accretion
rate, and various feedback efficiencies based on direct
simulations of inflows and outflows in AGN, on sub-parsec and parsec
scales performed by \citet{Kurosawa:2009b} (KP09 hereafter). In other words,
we wish to determine if AGN 
can supply energy in the form and amount required by
the cosmological and galaxy merger simulations.
\textcolor{black}{
In this work, we do not attempt to provide a definitive answer to the
problem of AGN feedback efficiency as our simulations do not include
the smallest scales (the black hole radius) and the very large scales
($>10$~pc), and also do not include all physical processes operating
in AGN (e.g., dust, magnetic fields and star formation).  Here, we
simply report the AGN feedback efficiencies found in the simulations
previously presented by KP09 who focused on relatively large scale
inflow and outflow ( $\sim0.01$ to $\sim10$~pc).
}

\section{AGN Models in Current Cosmological Simulations}

\label{sec:AGN-SPH}

\subsection{Mass Accretion Rates}

\label{sub:Mdot-SPH}

In recent cosmological and galaxy merger simulations (e.g., SDH05;
\citealt{Robertson:2006}; \citealt{Sijacki:2007}; \citealt{Khalatyan:2008};
\citealt{DiMatteo:2008}; \citealt{Johansson:2009}; BS09),
the actual physical process of the mass-accretion onto the BH is not
explicitly modeled because of a relatively poor resolution. Often,
these simulations rely on a separate analytical model to describe
the small scale physical processes. The unresolved accretion process
is usually described by a Bondi-Hoyle-Little formulation (\citealt{Hoyle:1939};
\citealt{Bondi:1944}; \citealt{Bondi:1952}). Here, we consider the
case in which the accreting BH does not move with respect to the surrounding
gas. This reduces the process to a simpler Bondi spherical accretion
problem. Then, the mass-accretion rate can be written as  
Equation~(\ref{eq:mdot_bondi}). 
The dimensionless constant $\lambda$ (in Eq.~[\ref{eq:mdot_bondi}])
depends on the adiabatic index $\gamma$.  For the gas with
$\gamma=5/3,$ $\lambda=1/4$ (see e.g., \citealt{Frank:1992}). The Bondi
accretion formula relates the mass-accretion rate of a BH located at
the center to the gas density and the sound speed (or equivalently the
temperature) of the gas at a large scale.

On the other hand, in the Bondi accretion prescription
used by e.g., SDH05 and BS09, the mass-accretion
rate is written as 
\begin{equation}
  \dot{M_{\mathrm{S}}}=\alpha\,\frac{4\pi
    G^{2}M_{\mathrm{BH}}^{2}\rho}{c_{\mathrm{s}}^{3}}
\label{eq:mdot_bondi_sph}
\end{equation}
where $\rho$ and $c_{\mathrm{s}}$ are the density and the sound
speed estimated near the BH using the surrounding smoothed particle
hydrodynamics (SPH: \citealt{Gingold:1977}; \citealt{Lucy:1977}) gas
particles. Note that the expression contains {}``the dimensionless
parameter'' $\alpha$ which is different from $\lambda$ in
Equation~(\ref{eq:mdot_bondi}). 
SDH05 introduced $\alpha$ parameter
to overcome the gap in the scale sizes between the numerical resolution
and the Bondi accretion regime. In a typical cosmological or galaxy
merger SPH simulation, the smoothing length ($\sim10^{3}$~pc) is much larger
than the gravitational radius of influence or the Bondi radius, 
$r_{\mathrm{B}}$, which is $\sim2$~pc.  
If we assume the gas located at a large distance,
heated by the AGN radiation, is {}``Comptonized''
($T\approx2\times10^{7}$~K), the corresponding speed of sound
(assuming $\gamma=5/3$) is 
relatively high ($\sim500\,\mathrm{km\, s^{-1}}$). 

SDH05, BS09 and others (e.g., \citealt{Robertson:2006};
\citealt{Sijacki:2007}; \citealt{Khalatyan:2008}; \citealt{DiMatteo:2008};
\citealt{Johansson:2009}) find that a very large factor of $\alpha$
is required for low-mass BHs to grow their masses; hence, the problem
is not strictly a Bondi accretion problem. Most of the AGN feedback
model in the cosmological simulations mentioned above assume a constant
value of $\alpha=100$ (see also Table~2 in BS09) except for BS09
who allow $\alpha$ to depend on the value of local gas density.
The assumption of a very large value of $\alpha$ becomes inadequate
when the local gas density is higher than that required by the formation
of the a cold interstellar gas phase, and when the cosmological simulation
does resolve the Jean length and the Bondi radius (BS09). 

By noting these, BS09 abandoned the assumption of constant $\alpha$ in
Equation~(\ref{eq:mdot_bondi_sph}), 
and introduced the following parametrization of $\alpha$. 
\begin{equation}
  \alpha=
  \begin{cases}
    1 & \mathrm{for\, n_{\mathrm{H}}<n_{\mathrm{H}}^{*}}\\
    \left(n_{\mathrm{H}}/\mathrm{n_{\mathrm{H}}^{*}}\right)^{\beta} &
    \mathrm{for}\, n_{\mathrm{H}}\geq n_{\mathrm{H}}^{*}
  \end{cases}
  \label{eq:alpha-booth}
\end{equation}
 where $n_{\mathrm{H}}$ and $n_{\mathrm{H}}^{*}$ are the number
density of hydrogen and the critical hydrogen number density above
which the gas is expected to become multi-phase, and star formation is
expected to begin via contraction of gas due to thermo-gravitational
instability (cf.~\citealt{Schaye:2004}; \citealt{Schaye:2008}).
The critical density is chosen as $n_{\mathrm{H}}^{*}=0.1\,\mathrm{cm^{-3}}$,
 i.e., the corresponding critical hydrogen density is
 $\rho_{\mathrm{H}}^{*}=1.7\times10^{-25}\,\mathrm{g\, cm^{-3}}$. 
The best fit models of BS09 to some observations (e.g., the $M_{\mathrm{BH}}$--$\sigma$
relation) gives $\beta=2.0$. Note that the new parametrization of
$\alpha$ in Equation~(\ref{eq:alpha-booth}) provides an additional
density dependency of the mass-accretion rate in Equation~(\ref{eq:mdot_bondi_sph}).
In the formulation of BS09, the mass-accretion rate steeply depends
on the density of the surrounding gas i.e., $\dot{M}\propto\rho^{3}$,
for $\rho>\rho_{\mathrm{H}}^{*}$ while the formulation of SDH05
and the original Bondi accretion model (Eq.~{[}\ref{eq:mdot_bondi}]) always
give a linear dependency i.e., $\dot{M}\propto\rho$. In most of the
AGN accretion models in the cosmological simulations (e.g., SDH05;
BS09), the highest mass-accretion rate is limited to the Eddington
rate, i.e., 
\begin{equation}
  \dot{M}_{\mathrm{Edd}}=\frac{4\pi
    GM_{\mathrm{BH}}m_{\mathrm{p}}}{\epsilon_{\mathrm{r}}\sigma_{\mathrm{T}}c}
\label{eq:mdot-Eddington}
\end{equation}
 where $m_{\mathrm{p}}$, $\epsilon_{\mathrm{r}}$, $\sigma_{\mathrm{T}}$
and $c$ are the proton mass, the radiative efficiency (the rest mass
to radiation conversion efficiency), the Thomson cross-section and
the speed of light, respectively. The Eddington ratio ($\Gamma$) is
defined as the ratio of a system mass-accretion rate to the Eddington
rate, i.e., $\Gamma = \MDOT_{\mathrm{a}} / \dot{M}_{\mathrm{Edd}}$.

\begin{figure}
\begin{center}

\includegraphics[clip,width=0.45\textwidth]{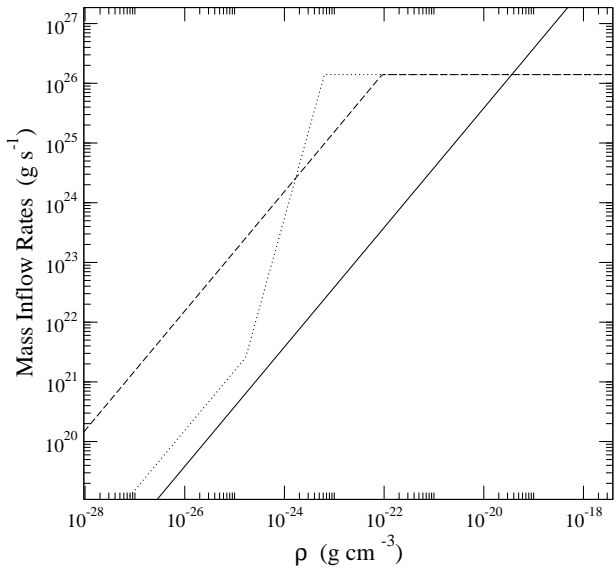}

\end{center}

\caption{Comparison of the mass-accretion rates from the Bondi accretion model
(\citealt{Bondi:1952}) (\emph{solid}), the modified Bondi accretion
model by SDH05 (\emph{dashed}) and that by BS09
(\emph{dotted}), as described in Eqs.~(\ref{eq:mdot_bondi}), (\ref{eq:mdot_bondi_sph})
and (\ref{eq:alpha-booth}), respectively. The mass-accretion rates
are computed as a function of density $\rho$ at a radius $r_{\mathrm{o}}$
which is much larger than the Bondi radius ($r_{\mathrm{B}}$), i.e.,
$r_{\mathrm{o}}\gg r_{\mathrm{B}}$. In all models, the speed of sound
$c_{\mathrm{s}}$ at $r_{\mathrm{o}}$ is set to $520\,\mathrm{km\, s^{-2}}$,
which corresponds to that of Comptonized gas temperature $T\approx2\times10^{7}$~K
with the adiabatic index of gas $\gamma=5/3$. The mass of the BH
is set to $M_{\mathrm{BH}}=10^{8}\,\Msun$. While $\alpha=100$ is
adopted in the model of SDH05, $\beta=2.0$ is adopted
in the model of BS09. In the modified Bondi accretion
models of SDH05 and BS09, the mass-accretion
rates are limited to the Eddington rate (Eq.~{[}\ref{eq:mdot-Eddington}]). }

\label{fig:mdot-models}
\end{figure}

Figure~\ref{fig:mdot-models} illustrates how the mass-accretion rate
depends on the density ($\rho$) in 
the models by \citet{Bondi:1952}, SDH05 and BS09. 
The mass of the BH is assumed as
$M_{\mathrm{BH}}=10^{8}\,\Msun$. 
In all three models, the speed of sound $c_{\mathrm{s}}$ is set to
$520\,\mathrm{km\, s^{-2}}$, which corresponds to that of Comptonized
gas temperature $T\approx2\times10^{7}$~K with the adiabatic index
of gas $\gamma=5/3$. In the modified Bondi accretion models of SDH05
and BS09, the mass-accretion rates are limited to the Eddington rate
(Eq.~{[}\ref{eq:mdot-Eddington}]), and the radiative efficiency
$\epsilon_{\mathrm{f}}$ is set to $0.1$. The dimensionless parameter
$\alpha=100$ is adopted for the model of SDH05, and $\beta=2.0$
is adopted in the model of BS09. The figure shows that the mass-accretion
rate of SDH05 is larger than the Bondi accretion rate for $\rho\simless10^{-20}\,\mathrm{g\, cm^{-3}}$.
On the other hand, the mass-accretion rate by BS09 is similar to
that of the Bondi accretion rate (off by a factor of $\lambda=1/4$
in Eq.~{[}\ref{eq:mdot_bondi}]) in the low-density regime ( $\rho\simless10^{-25}\,\mathrm{g\, cm^{-3}}$),
but it is significantly larger than the Bondi accretion rate for the
density range of $10^{-25}\simless\rho\simless10^{-20}\,\mathrm{g\, cm^{-3}}$.
The densities above which the accretion proceeds at the Eddington rate
are $\sim9.5\times10^{-23}\,\mathrm{g\, cm^{-3}}$ and $\sim6.3\times10^{-24}\,\mathrm{g\, cm^{-3}}$
for the models of SDH05 and BS09, respectively. However, these values
change depending on the adopted value of $c_{\mathrm{s}}$. In the
Bondi accretion model, the mass-accretion rates reaches the Eddington
rate at much higher density ($\rho\sim10^{-20}\,\mathrm{g\, cm^{-3}}$). 
We note that adding a rotation to the gas can reduce the mass-accretion
rate. As shown by \citet{Proga:2003c} and \citet{Ryu:1995}, the
reduction of $\MDOT_{\mathrm{a}}$ can be significant, i.e., by one
order of magnitude or more, compared to $\MDOT_{\mathrm{B}}$





\subsection{AGN Feedback Models}

\label{sub:AGN-FB-SPH}

In SDH05 and BS09 (also in many others, cf.~Table~2 in BS09),  $\epsilon_{\mathrm{r}}=0.1$
(e.g., \citealt{shakura:1973}; see also \citealt{Soltan:1982})
is assumed, and is fixed. A higher value of $\epsilon_{\mathrm{r}}$
($\sim0.2$) can be achieved in an accretion model with a thin disk
and a rapidly rotating BH (e.g., \citealt{Thorne:1974}).
Recent observations suggest a wide range of $\epsilon_{\mathrm{r}}$:
$0.07$ (\citealt{Martinez-Sansigre:2009}), $0.30$--$0.35$
(\citealt{Wang:2006b}), $0.16$ (\citealt{Yu:2008}), $0.15$ (\citealt{Elvis:2002})
 and $\sim 0.1$ or $\sim 0.2$ (\citealt{Yu:2002}). 
On the other hand, \citet{Cao:2008} find $\epsilon_{\mathrm{r}}$
is relatively low ($\sim0.08$) for $M_{\mathrm{BH}}<10^{8}\,\Msun$
and relatively high ($\simgreat0.18$) for $M_{\mathrm{BH}}\simgreat10^{9}\,\Msun$.
Since the exact mechanism of how the accretion luminosity of a BH
couples to the surrounding gas is not well known, 
SDH05 and BS09 simply assume that $L_{\mathrm{a}}$ couples only
thermally (and isotropically) to the surrounding. Using
Equation~(\ref{eq:L-acc}), the rate of energy deposition to the
surrounding (the AGN feedback rate) in SDH05 is written as 
\begin{equation}
  \dot{E}_{\mathrm{f}}=\epsilon_{\mathrm{f}}\,
  L_{\mathrm{a}}=\epsilon_{\mathrm{f}}\epsilon_{\mathrm{r}}\dot{M}_{\mathrm{BH}}c^{2}
\label{eq:Edot-Feedback}
\end{equation}
 where $\epsilon_{\mathrm{f}}$ is the efficiency of the AGN energy
deposition to the surrounding gas, and is a free parameter which is to
be constrained by observations. BS09 find the models
with $\epsilon_{\mathrm{f}}=0.15$ matches observations (e.g., the Magorrian
relation and the $M_{\mathrm{BH}}-\sigma$ relations) very well, and
similarly SDH05 find $\epsilon_{\mathrm{f}}=0.05$ matches
observations well (see also \citealt{DiMatteo:2005}). In the study
of BS09, they find that the global BH number density at the current
era (zero redshift) and the BH scaling relations are very sensitive
to a choice of $\epsilon_{\mathrm{f}}$, and the former is nearly inversely
proportional to the value of $\epsilon_{\mathrm{f}}$.

\section{Our Model}

\label{sec:Model}

Our approach is to use physical two-dimensional (axisymmetric) and time-dependent
hydrodynamical (HD) simulations of AGN flows to investigate the dependency
of the BH mass-accretion rate on the surrounding gas density, and to
find the AGN feedback efficiencies in converting the accretion
luminosity into the outward fluxes of energy, momentum and mass. 
Here, we simply analyze the simulations results previously published in KP09
for this purpose.  KP09 used a modified version of the {\sc ZEUS-MP} code
\citep[cf.][]{Hayes:2006} for their numerical simulations.
In the following, we briefly summarize their main
model assumptions and results.

In KP09, we studied axisymmetric hydrodynamical simulations of a slowly 
rotating gas that is under the influence of the gravity of 
a $10^8~\MSUN$ black hole and is irradiated by 
a geometrically thin UV accretion disk and a spherical X-ray corona. 
\textcolor{black}{
   We assumed that the AGN radiation is dominated by the disk radiation
   (95\% of the total luminosity). Further, we account for the fact
   that the radiation
   from the disk depends on the polar angle $\theta$, i.e., proportional
   to $\cos{\theta}$ due to the geometrical foreshortening.
}
We ran a set of simulations for various values of 
the gas density ($\rho_{\mathrm{o}}$) at the outer
radius of the computational domain, $r_{\mathrm{o}}\approx$~7 pc. After the
initial transient stage,  this density determines the key
characteristics of our solutions such as the accretion luminosity and
the outflow properties. We compute the accretion luminosity of a system
based on the accretion-rate which is assumed to be equal to 
the mass-supply rate at the inner radius of the computational domain 
$r_{\mathrm{i}}\approx 10^{-2}$ pc, (i.e., we used
Eq.~{[}\ref{eq:L-acc}] where we assumed $\epsilon_{\mathrm{r}}=1/12$
and  $\MDOT_{\mathrm{a}}=\MDOT_{\mathrm{in}}[r_{\mathrm{i}}]$).
\textcolor{black}{
See \S~\ref{subsec:Mdot_a_explain} for more detail on this
assumption. 
}

For the models with high temperature gas at large radii and with high 
luminosities, we found a strong correlation between  $\MDOT_{\mathrm{out}}$  and
$L_{\mathrm{a}}$ (see Fig.~1 in KP09).  The power law index
describing the correlation is very similar to that for
radiation-driven stellar and disk wind models (e.g.,
\citealt{Castor:1975};  \citealt{Proga:1998};
\citealt{Proga:1999}). More surprisingly, for the models with high
density at large radii, we found  that steady state solutions with  
$L_{\mathrm a}$ exceeding the Eddington limit. The super-Eddington accretion 
proceeds in the equatorial region and is possible because the radiation flux 
from the disk is significantly reduced in the equatorial direction due to 
the geometrical foreshortening effect. 

\begin{figure*}

\begin{center}
\includegraphics[clip,width=0.98\textwidth]{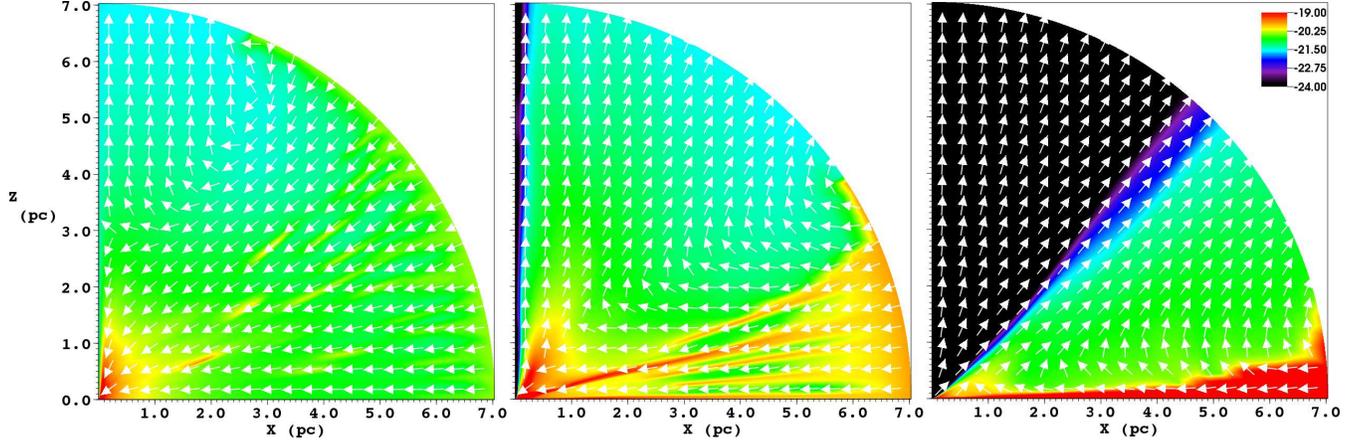}
\par\end{center}

\caption{Examples of the density and velocity maps from the set of
simulations (without the  temperature constrained at the outer boundary) performed by  \citet{Kurosawa:2009b}. The density (in logarithmic scale)
  is over-plotted with the directions of poloidal velocity as arrows
  for Models~36 (\emph{left panel}), 29 (\emph{middle panel}) and 34
  (\emph{right panel}). See Table~\ref{tab:Model-Summary} 
for the corresponding model  numbers. The figures are placed in order of 
increasing density at the outer boundaries ($\rho_{\mathrm{o}}$) from 
the left to right  (cf.~Table~\ref{tab:Model-Summary}).  
Their corresponding accretion luminosities in the unit of 
the Eddington luminosity, i.e., the Eddington ratio, are 
$\Gamma = 0.32, 0.71$ and $4.3$, from the  left to right panels. 
  For the low density ($\rho_{\mathrm{o}}$) and low accretion
  luminosity  model (Model~36 in the left panel), the outflow is  
very narrow ($0^{\circ} \leq \theta  \simless30^{\circ}$), 
and the inflow is very wide
($30^{\circ} \simless \theta \simless 90^{\circ}$).  As the
  density ($\rho_{\mathrm{o}}$) and accretion luminosity increase
  (Model~29 in the middle panel), the outflow becomes wider
  ($0^{\circ} \leq  \theta \simless 50^{\circ}$), whereas the inflow
  becomes narrower ($50^{\circ} \simless \theta \simless 90^{\circ}$). For
  the very high density and accretion model (Model~34 in the right
  panel), the outflow occurs over a  very wide range of the polar angle 
($0^{\circ} \leq
  \theta \simless 85^{\circ}$), and the accretion region is now confined to
  a thin equatorial wedge (the disk-wind-like solution).  }

\label{fig:samples}

\end{figure*}

In all simulations performed by KP09, an outflow is driven
from an inflow with sub-Keplerian rotation. For the models with high
temperatures at large radii, the inflow occurs over a wide range of the polar angles, 
whereas the outflow occurs in a relatively narrow polar angle range
(see the left panel in Fig.~\ref{fig:samples}). However, 
for the super-Eddington cases with low temperature at large radii, 
the inflow persists only very close to the equatorial plane, 
resembling a thin accretion disk, 
while the outflow arises in a wide range of radii and polar angles 
(see the right panel in Fig.~\ref{fig:samples}). 
The geometry of this extreme inflow-outflow solution is very similar to 
a radiation-driven wind from a luminous Keplerian accretion disk
(e.g., \citealt{Woods:1996}; \citealt{Proga:1998}; \citealt{Proga:2002b}).
For the cold super-Eddington solutions, $\MDOT_{\mathrm{out}}$ is only very 
weakly correlated. The weaker correlation is mainly caused 
by a mismatch between with the direction of escaping photons 
and the inflowing gas. In other words, the radiation is emitted mostly
in the polar directions whereas the inflowing gas occurs mainly in
the equatorial region. 

As it has been discussed and shown in 
the past, we find that self-consistently determined preheating/cooling from 
the quasar radiation can significantly reduce the rate at which 
the central BH is fed with matter. However, our results also emphasize 
a little-appreciated feature. Namely, quasar radiation does drive a non-spherical, 
multi-temperature and very dynamic flow. 

In the following, we present the mass-accretion rates and
various (energy, momentum and mass) AGN feedback efficiencies computed
from the simulations. For this purpose, we use a subset of the models
in KP09. Here, we concentrate on the models in which \emph{the outer
boundary temperature is not fixed at a constant value, but it is
self-consistently determined from the radiative and adiabatic heating}
(Models~28--34 in KP09).  Note that the flow solutions for these models
are, in general, very similar to those with the fixed outer boundary
temperature at $2\times10^6\,\Kelvin$ (the low temperature models,
i.e., Models~1-9 in KP09).




\begin{table*}

\caption{Model Summary}

\label{tab:Model-Summary}

\begin{center}


\begin{tabular}{rcccccc}
 &  &  &  &  &  & \tabularnewline
\hline
\hline 
 & $\rho_{\mathrm{o}}$ & $T_{\mathrm{o}}^{**}$ & $\dot{M}_{\mathrm{in}}\left(r_{\mathrm{i}}\right)$ & $\Gamma$ & $\dot{M}_{\mathrm{out}}\left(r_{\mathrm{o}}\right)$ & $\dot{M}_{\mathrm{in}}\left(r_{\mathrm{o}}\right)$\tabularnewline
Model$^{*}$ & $\left(10^{-21}\,\mathrm{g\, cm^{-3}}\right)$ & $\left(10^{7}\,\mathrm{K}\right)$ & $\left(10^{25}\mathrm{\, g\, s^{-1}}\right)$ & $\cdots$ & $\left(10^{25}\,\mathrm{g\, s^{-1}}\right)$ & $\left(10^{25}\,\mathrm{g\, s^{-1}}\right)$\tabularnewline
\hline
35 & 2 & 0.02 & $3.4\left(0.1\right)^{\dagger}$ & $0.20\left(0.01\right)$ & $0.0\left(0.0\right)$ & $3.4\left(0.1\right)$\tabularnewline
36 & 4 & 0.07 & $5.5\left(0.6\right)$ & $0.32\left(0.04\right)$ & $0.11\left(0.02\right)$ & $5.4\left(0.2\right)$\tabularnewline
28 & 10 & 0.14 & $8.9\left(0.68\right)$ & $0.52\left(0.032\right)$ & $1.0\left(0.032\right)$ & $9.4\left(0.1\right)$\tabularnewline
29 & 20 & 0.14 & $12\left(0.58\right)$ & $0.71\left(0.034\right)$ & $3.5\left(0.52\right)$ & $15\left(0.3\right)$\tabularnewline
30 & 40 & 0.36 & $18\left(1.1\right)$ & $1.1\left(0.1\right)$ & $7.1\left(1.4\right)$ & $25\left(0.2\right)$\tabularnewline
31 & 80 & 0.80 & $25\left(2.6\right)$ & $1.4\left(0.2\right)$ & $10\left(2.3\right)$ & $35\left(2.1\right)$\tabularnewline
32 & 160 & 0.98 & $36\left(4.9\right)$ & $2.1\left(0.3\right)$ & $9.9\left(2.6\right)$ & $49\left(5.3\right)$\tabularnewline
33 & 320 & 0.85 & $52\left(2.6\right)$ & $3.1\left(0.2\right)$ & $11\left(0.48\right)$ & $63\left(1.1\right)$\tabularnewline
34 & 640 & 1.30 & $72\left(0.56\right)$ & $4.3\left(0.03\right)$ & $9.5\left(0.19\right)$ & $82\left(0.9\right)$\tabularnewline
\hline
\end{tabular}

\raggedright

\vspace{0.5cm}

({*})~The model numbers are identical to those in \citet{Kurosawa:2009b}
except for Models~35 and 36 which are additional models presented
here for the first time. 

({*}{*})~Self-consistently determined temperature at the outer boundary. 

($\dagger$)~Values in brackets are the standard deviations of the
time series values. 

\end{center}

\end{table*}

\section{Results}

\label{sec:Results}

We analyze the dependency of the mass-accretion rate on the gas
density at a large distance from a BH and AGN feedback efficiencies
based on the axisymmetric hydrodynamical simulations presented in
KP09.  The main results of the models along with the
input outer boundary density $\rho_{\mathrm{o}}$ are summarized in
Table~\ref{tab:Model-Summary}.

\subsection{Mass-Accretion Rates}

\label{sub:result-mdot}

\begin{figure}
\begin{center}

\includegraphics[clip,width=0.45\textwidth]{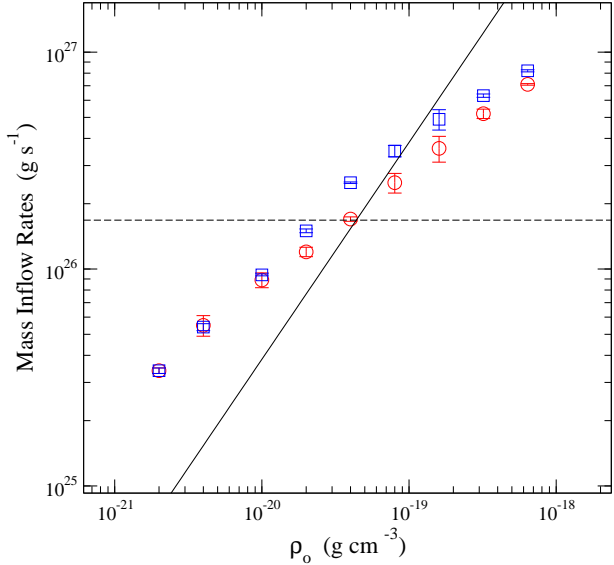}

\end{center}

\caption{Comparison of the mass-inflow rates found in the HD simulations 
(Tab.~\ref{tab:Model-Summary})
with those predicted by the Bondi accretion model (\citealt{Bondi:1952})
(\emph{solid line}) and with those adopted by SDH05
and BS09 (\emph{dashed line}). In the density range
of the models considered here, the mass-accretion rates adopted by
SDH05 and BS09 are limited by the
Eddington rate (Eq.~{[}\ref{eq:mdot-Eddington}]); hence, the line
is flat (cf.~Fig.~\ref{fig:mdot-models}). The mass-inflow rates
at the inner boundary (\emph{circles}) and those at the outer boundary
(\emph{squares}) of the computational domain are shown as a function
of the outer boundary density $\rho_{\mathrm{o}}$. The mass-accretion
from the HD simulations are very similar to the Bondi accretion rates,
but the HD models has a less steeper dependency on the density. The
Bondi mass-accretions rates and those of SDH05 and
BS09 are computed for the gas with the Comptonized temperature
$T=2\times10^{7}$~K and with the adiabatic index $\gamma=5/3$. }

\label{fig:mdot}
\end{figure}

The mass-inflow rates at the inner boundary $\dot{M}_{\mathrm{in}}\left(r_{\mathrm{i}}\right)$
and those at outer boundary $\dot{M}_{\mathrm{in}}\left(r_{\mathrm{o}}\right)$
from the HD simulations are plotted as a function of the outer boundary
density $\rho_{\mathrm{o}}$ in Figure~\ref{fig:mdot} (see Tab.~\ref{tab:Model-Summary}
for the numerical values). For a given value of $\rho_{\mathrm{o}}$,
$\dot{M}_{\mathrm{in}}\left(r_{\mathrm{i}}\right)$
and $\dot{M}_{\mathrm{in}}\left(r_{\mathrm{o}}\right)$ are not equal
to each other, but rather
$\dot{M}_{\mathrm{in}}\left(r_{\mathrm{i}}\right)<\dot{M}_{\mathrm{in}}\left(r_{\mathrm{o}}\right)$
because of an outflow. The lowest density model (Model~35) is an exception
since no outflow is formed in this model. For the higher density models,
an outflow forms, and not all the material entering from the outer
boundary reaches the inner boundary. A fraction of gas experiences
a strong radiation pressure and radiative heating, and the direction
of flow changes, forming an outflow.  

The figure also shows the the mass-accretion rates predicted by the
Bondi accretion model (Eq.~{[}\ref{eq:mdot_bondi}]) and those computed
from the formulations of SDH05 and BS09 (Eqs.~{[}\ref{eq:mdot_bondi_sph}]
and {[}\ref{eq:alpha-booth}]). The outer radius
is much smaller than that of a typical smoothing scale on a SPH cosmological
simulation ($\sim10^{3}$~pc), and the outer density values used
in our simulations are much larger than a typical local density at
BH in the SPH simulations. In our simulations, the higher density
at a 10~pc scale is required for a system to produce an outflow.
For example, as we can see in Table~\ref{tab:Model-Summary}, $\rho_{\mathrm{o}}$
must be greater than $2\times10^{-21}\,\mathrm{g\, cm^{-3}}$, which
corresponds to ($n_{\mathrm{H}}\simgreat1.2\times10^{3}\,\mathrm{cm^{-3}}$),
to form an outflow with our system setup. In the density
range of the models considered here, the mass-accretion rates adopted
by SDH05 and BS09 are limited by the Eddington rate (Eq.~{[}\ref{eq:mdot-Eddington}]);
hence, the line is flat (cf.~Fig.~\ref{fig:mdot-models}). Note
that the radiative efficiency $\epsilon_{\mathrm{r}}=1/12$ 
instead of 0.1 is adopted for the Eddington rate (Eq.~{[}\ref{eq:mdot-Eddington}])
in the models of SDH05 and BS05 to be consistent with our simulations.
This moves the Eddington rate only slightly upward.

Our models include the effects of radiative heating and radiation
force. Therefore, we do not, in general, expect our solution to reproduce 
an exactly
same density dependency of the mass-inflow rate as that of the Bondi
model. However, the figure shows the mass-inflow rates from our models
are very similar to those of the Bondi rates, i.e., the rates are
of the same order of magnitude. Interestingly our $\dot{M}_{\mathrm{in}}\left(r_{\mathrm{i}}\right)$
and the Bondi mass-accretion rate matches around $\rho_{\mathrm{o}}=4\times10^{-20}\,\mathrm{g\, cm^{-3}}$
which coincidentally corresponds to $\Gamma\approx1$. Since the accretion
rates from SDH05 and BS09 are the Eddington rates (the rates corresponding
to $\Gamma=1$) in this density range, their lines also cross at the same point. 

The figure clearly shows that our models have a weaker dependency of the
mass-inflow rates on the density than that of the Bondi accretion. The power-law fits of data points
give the slope $q=0.52\left(\pm0.01\right)$ for $\dot{M}_{\mathrm{in}}\left(r_{\mathrm{i}}\right)$
and $q=0.56\left(\pm0.02\right)$ for $\dot{M}_{\mathrm{in}}\left(r_{\mathrm{o}}\right)$,
which are indeed much smaller than that of the Bondi accretion model,
i.e., $q=1$ (cf.~Eq.~{[}\ref{eq:mdot_bondi}]). 
Although not shown here, if we turn off the rotation, radiation force and radiative heating in
our models, we obtain $q\approx1$, as this is equivalent to the Bondi
accretion problem (see also \citealt{Proga:2003c}; \citealt{Janiuk:2008}).


\subsection{Feedback Efficiencies}

\label{sub:efficiency-HD}

Next, we compute AGN feedback efficiencies in energy, momentum and
mass using the simulation results, as defined in
Equations~(\ref{eq:eff-total-energy})--(\ref{eq:eff-mass}). Since the models
used here show some degree of variability (typically $\sim10\%$ level,
cf.~Tab~\ref{tab:Model-Summary}), the physical quantities used to
compute the feedback efficiencies are based on the time averaged
values.


\begin{table*}

\caption{Feedback Efficiencies}

\label{tab:efficiency}

\begin{center}

\vspace{-0.5cm}

\begin{tabular}{ccccccc}
 &  &  &  &  &  & \tabularnewline
\hline
\hline 
Model & $\Gamma$ & $\epsilon_{\mathrm{k}}$ & $\epsilon_{\mathrm{th}}$ & $\epsilon_{\mathrm{t}}$ & $\epsilon_{\mathrm{p}}$ & $\epsilon_{\mathrm{m}}$\tabularnewline
\hline
35 & $0.20$ & $0$ & $0$ & $0$ & $0$ & $0$\tabularnewline
36 & $0.32$ & $1.4\times10^{-8}$ & $9.1\times10^{-8}$ & $1.0\times10^{-7}$ & $7.8\times10^{-5}$ & 0.02\tabularnewline
28 & $0.52$ & $8.1\times10^{-7}$ & $4.6\times10^{-7}$ & $1.3\times10^{-6}$ & $1.3\times10^{-3}$ & 0.11\tabularnewline
29 & $0.71$ & $1.4\times10^{-5}$ & $1.0\times10^{-6}$ & $1.5\times10^{-5}$ & $8.3\times10^{-3}$ & 0.29\tabularnewline
30 & $1.1$ & $9.0\times10^{-5}$ & $1.8\times10^{-6}$ & $9.2\times10^{-5}$ & $2.9\times10^{-2}$ & 0.39\tabularnewline
31 & $1.4$ & $1.3\times10^{-4}$ & $9.0\times10^{-6}$ & $1.4\times10^{-4}$ & $3.0\times10^{-2}$ & 0.40\tabularnewline
32 & $2.1$ & $1.0\times10^{-4}$ & $2.0\times10^{-6}$ & $1.0\times10^{-4}$ & $1.8\times10^{-2}$ & 0.28\tabularnewline
33 & $3.1$ & $5.8\times10^{-5}$ & $1.7\times10^{-6}$ & $6.0\times10^{-5}$ & $9.8\times10^{-3}$ & 0.21\tabularnewline
34 & $4.3$ & $8.2\times10^{-5}$ & $1.5\times10^{-6}$ & $8.4\times10^{-5}$ & $7.7\times10^{-3}$ & 0.13\tabularnewline
\hline
\end{tabular}

\end{center}

\end{table*}


\subsubsection{Energy Feedback Efficiency}

\label{subsub:eff_energy}

\begin{figure}
\begin{center}

\includegraphics[clip,width=0.45\textwidth]{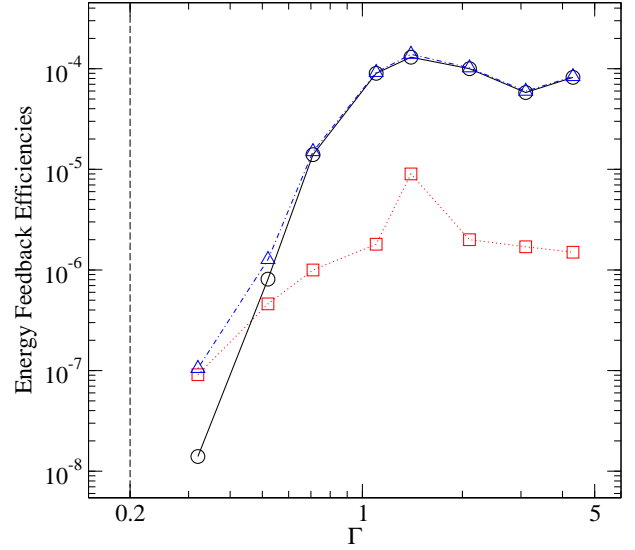}

\end{center}

\caption{The efficiencies of converting the BH accretion luminosity
  $L_{\mathrm{a}}$ to the rate of energy deposition to the surrounding
  gas are plotted as a function of the Eddington ratio
  ($\Gamma$). The panel shows the kinetic energy feedback efficiency
  $\epsilon_{\mathrm{k}}$ (\emph{circles}), the thermal energy
  feedback efficiency $\epsilon_{\mathrm{th}}$ (\emph{squares}) and
  the total energy feedback efficiency
  $\epsilon_{\mathrm{t}}=\epsilon_{\mathrm{k}}+\epsilon_{\mathrm{th}}$
  (\emph{triangles}), separately (see
  Eqs.~{[}\ref{eq:eff-total-energy}], {[}\ref{eq:eff-kinetic-energy}],
  and {[}\ref{eq:eff-thermal-energy}]).  The maximum total energy
  feedback efficiency is $\sim10^{-4}$. 
  For the models with relatively low Eddington
  ratio ($\Gamma$$\simless0.4$), the thermal feedback is more
  efficient than the kinetic feedback
  ($\epsilon_{\mathrm{th}}>\epsilon_{\mathrm{k}}$).  For the models with
  relatively high Eddington ratio ($\Gamma \simgreat0.6$), the kinetic
  feedback is more efficient than the thermal feedback by a factor of
  $\sim10$ to $\sim100$. The model with $\Gamma=0.2$ does not form an
  outflow, and the vertical line (\emph{dashed}) at $\Gamma=0.2$
  indicates an approximate $\Gamma$ value below which no outflow forms.
  The flattening of the efficiencies at $\Gamma\approx1$ is caused by
  the transition of the inflow-outflow morphology to a {}``disk wind''
  like configuration for the larger $\Gamma$ models (cf.~Fig.~\ref{fig:samples}).}

\label{fig:eff_energy}
\end{figure}

Figure~\ref{fig:eff_energy} shows the energy feedback efficiencies,
$\epsilon_{\mathrm{t}}$, $\epsilon_{\mathrm{k}}$ and
$\epsilon_{\mathrm{th}}$ computed based on our models
(Table~\ref{tab:Model-Summary}), as a function of the Eddington ratio
($\Gamma$). The numerical values of the efficiencies are listed in
Table~\ref{tab:efficiency}. For systems with relatively low Eddington
ratio ($\Gamma$$\simless0.4$), the thermal feedback efficiency is higher than the
kinetic feedback efficiency ($\epsilon_{\mathrm{th}}>\epsilon_{\mathrm{k}}$).  On
the other hand, for systems with relatively high Eddington ratio
($\Gamma$$\simgreat0.6$), the kinetic feedback dominates the thermal
feedback by a factor of $\sim10$ to $\sim100$. The model with
$\Gamma=0.2$ does not form an outflow, indicating an approximate
$\Gamma$ value below which no outflow forms (with our system setup).

The energy feedback efficiencies increase as $\Gamma$ increases, but
the efficiencies saturate for $\Gamma\simgreat1$. The total energy
feedback efficiency peaks at $\Gamma\approx1$ with
$\epsilon_{\mathrm{t}}\sim10^{-4}$.  The flattening of the
efficiencies for $\Gamma\simgreat1$ is caused by the transition of the
inflow-outflow morphology to a {}``disk wind'' like configuration
(cf.~Fig.~\ref{fig:samples}) for
the higher $\Gamma$ models (KP09). As briefly mentioned in
\S~\ref{sec:Model}, because of the mismatch between the
direction in which most of the radiation escapes (in polar direction)
and the direction in which the most of the accretion 
occurs in the system (the equatorial direction), the radiatively
driven outflows in the disk-wind-like configuration cannot increase
the outflow efficiency by increasing the accretion luminosity or
equivalently $\Gamma$. A similar behavior is found in the
$\dot{M}_{\mathrm{out}}\left(r_{\mathrm{o}}\right)$-- $\Gamma$ relation
of KP09 (see their Fig.~7).




\subsubsection{Momentum Feedback Efficiency}

\label{subsub:eff_momentum}

\begin{figure}
\begin{center}

\includegraphics[clip,width=0.45\textwidth]{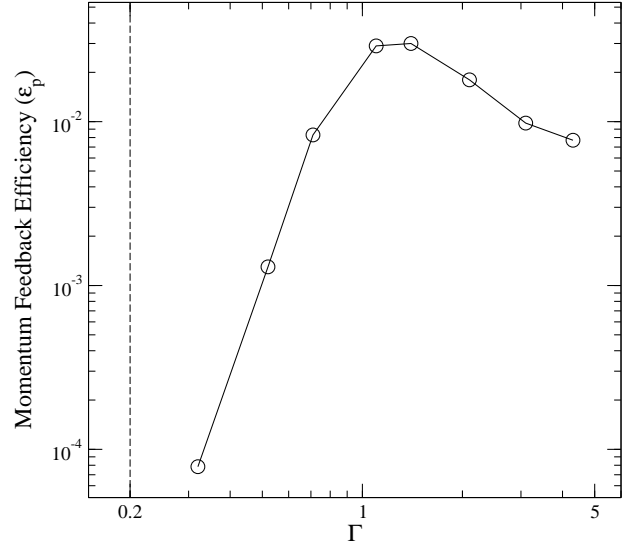}

\end{center}

\caption{The momentum feedback efficiency ($\epsilon_{\mathrm{p}}$), which
is defined as the ratios of the total wind momentum (at the outer
boundary) to the total radiation momentum ($L_{\mathrm{a}}/c$),
is plotted as a function of the Eddington ratio ($\Gamma$). The efficiency
peaks at $\Gamma\approx1$ with $\epsilon_{\mathrm{p}}\approx0.03$,
and it decreases for larger $\Gamma$ values. The decline of the curve
beyond $\Gamma\approx1$ is caused by the change in the inflow-outflow
morphology to a {}``disk wind'' like configuration for the larger
$\Gamma$ models (cf.~Fig.~\ref{fig:samples}). The model with $\Gamma=0.2$
does not form an outflow, and the vertical line (\emph{dashed}) at $\Gamma=0.2$
indicates an approximate $\Gamma$ value below which no outflow forms.}

\label{fig:eff_momentum}
\end{figure}

Figure~\ref{fig:eff_momentum} shows 
the momentum feedback efficiency $\epsilon_{\mathrm{p}}$ as a
function $\Gamma$.  The numerical values of
$\epsilon_{\mathrm{p}}$ for each model are listed in
Table~\ref{tab:efficiency}. The dependency of $\epsilon_{\mathrm{p}}$
on $\Gamma$ is similar to that of the energy feedback efficiency.  For
$\Gamma\simless1$, $\epsilon_{\mathrm{p}}$ increases as $\Gamma$
increases, but it decreases as $\Gamma$ increases for
$\Gamma\simgreat1$.  No outflow is formed for the models with
$\Gamma=0.2$ and below.  The cause of the peaking of
$\epsilon_{\mathrm{p}}$ at $\Gamma\approx1$ (and the declining for
$\Gamma\simgreat1$) is again due to the change in the inflow-outflow
morphology to a disk-wind-like (KP09; see also
Fig.~\ref{fig:samples}). The momentum deposition of the 
photons becomes less efficient once the flow has the
disk-wind-like configuration since a major fraction of the radiation
escapes in the polar direction, but gas is not present in that
direction since it has been already blown away by the strong
radiation. The maximum momentum feedback efficiency is
$\epsilon_{\mathrm{p}}\approx0.03$ which is about 2 orders of
magnitude larger than the total \emph{energy} feedback efficiency
found in \S~\ref{subsub:eff_energy}.




\subsubsection{Mass Feedback Efficiency}

\label{subsub:eff_mass}

\begin{figure}
\begin{center}

\includegraphics[clip,width=0.45\textwidth]{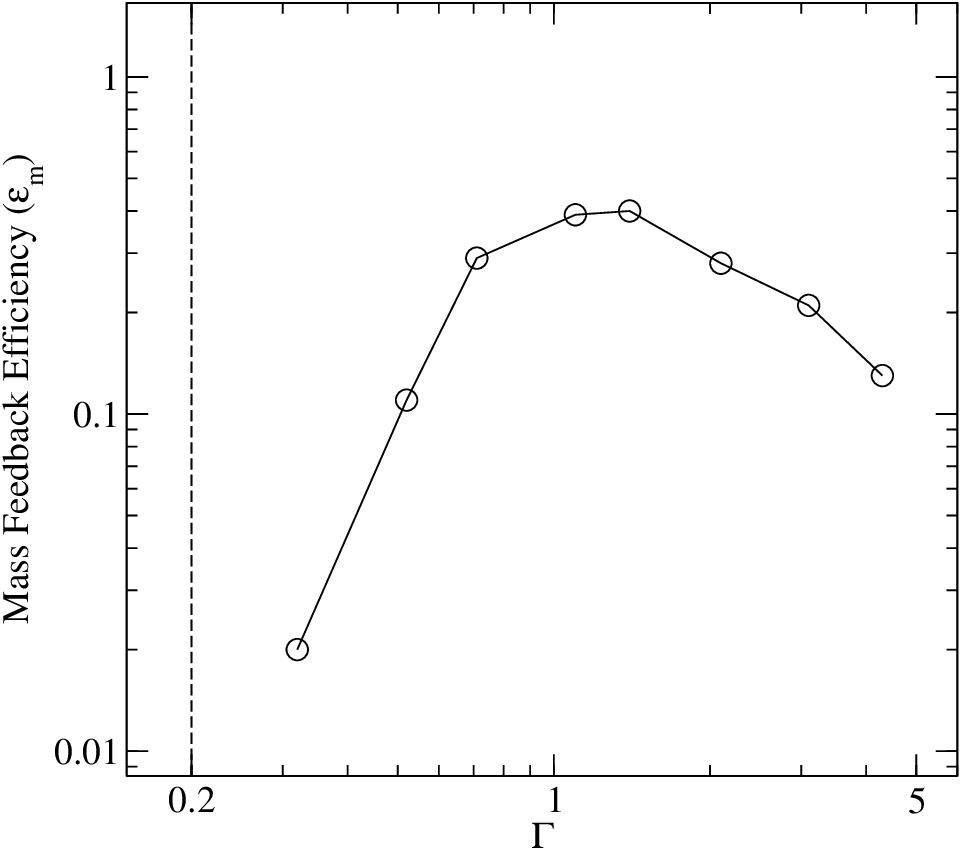}

\end{center}

\caption{The mass feedback efficiency ($\epsilon_{\mathrm{m}}$)
  plotted as a function the Eddington ratio ($\Gamma$). The efficiency
  $\epsilon_{\mathrm{m}}$ is defined as the ratio of the mass-outflow
  rate at the outer boundary
  $\dot{M}_{\mathrm{out}}\left(r_{\mathrm{o}}\right)$ to the
  mass-inflow rate at the inner boundary
  $\dot{M}_{\mathrm{in}}\left(r_{\mathrm{i}}\right)$, i.e.,
  $\epsilon_{\mathrm{m}}=\dot{M}_{\mathrm{out}}\left(r_{\mathrm{o}}\right)/\dot{M}_{\mathrm{in}}\left(r_{\mathrm{i}}\right)$.
  The efficiency peaks around $\Gamma=1$ with the maximum efficiency 
  value $\sim0.4$, i.e., $40\%$ of the total inflowing mass is
  converted to the outflows. The turn-around of
  $\epsilon_{\mathrm{m}}$ values is caused by the transition of the
  outflow morphology to a {}``disk wind'' like configuration for the
  larger $\Gamma$ models (cf.~Fig.~\ref{fig:samples}). The model
  with $\Gamma=0.2$ does not form  an outflow, and the vertical line 
  (\emph{dashed}) at $\Gamma=0.2$  indicates an approximate
  $\Gamma$ value below which no outflow forms.}
\label{fig:eff_mass}
\end{figure}

Figure~\ref{fig:eff_mass} shows the mass feedback efficiency ($\epsilon_{\mathrm{m}}$)
plotted as a function $\Gamma$. The dependency
of $\epsilon_{\mathrm{m}}$ on $\Gamma$ is very similar to that of
the momentum feedback efficiency (Fig.~\ref{fig:eff_momentum}).
The numerical values of $\epsilon_{\mathrm{m}}$ are listed in Table~\ref{tab:efficiency}.
For $\Gamma\simless1$, the mass feedback efficiency $\epsilon_{\mathrm{m}}$
increases as $\Gamma$ increases, but it starts to decreases slightly
beyond $\Gamma\approx1$. The efficiency peaks around $\Gamma=1$
with the maximum efficiency value $\sim0.4$. In other words, about
$40\%$ of the total inflowing mass is redirected
to an outflow. The turn-around of $\epsilon_{\mathrm{m}}$ values is
caused by the transition of the outflow morphology to a disk-wind-like
configuration (cf.~Fig.~\ref{fig:samples}) for the larger $\Gamma$
models, as in the cases for the energy and momentum feedback
efficiencies (\S\S~\ref{subsub:eff_energy} and \ref{subsub:eff_momentum}). 

The accretion process in our model is fundamentally different from
that of the Bondi accretion model and those adopted in the cosmological
simulations (e.g., SDH05; BS09) because in our model
an outflow and an inflow can simultaneously be formed while the Bondi
accretion model can only form an inflow. The model of SDH05 and others
can form either an inflow or an accretion (but not both simultaneously).




\section{Discussion}

\label{sec:Discussion}

\subsection{\textcolor{black}{Comparisons with Other Studies}}
\label{subsec:compare-others}

A main purpose of this study is to measure the efficiency of the AGN feedback
in our radiation  hydrodynamical simulations of accretion flows irradiated 
by AGN. 
Especially, we are interested in the energy feedback efficiency
($\epsilon_{\mathrm{f}}$) as this is an important parameter to determine
the BH growth rate and the BH number density at the current epoch
(redshift zero) in the cosmological simulations (e.g., SDH05; BS09).
BS09 find that the BH growth rate is nearly inversely proportional
to the value of $\epsilon_{\mathrm{f}}$, which is important for a
self-regulation of a BH growth (e.g., \citealt{Silk:1998};
\citealt{Fabian:1999}).

The approximate ratios of the maximum feedback efficiencies in energy,
momentum and mass using the results from \S~\ref{sub:efficiency-HD}
are: 
\begin{equation}
  \epsilon_{\mathrm{m}}:\epsilon_{\mathrm{p}}:\epsilon_{\mathrm{t}}:\epsilon_{\mathrm{k}}:\epsilon_{\mathrm{th}}=1000:100:1:1:0.1
  \label{eq:eff-ratios}
\end{equation}
 where the total energy feedback efficiency $\epsilon_{\mathrm{t}}\sim10^{-4}$.
Compared to the mass feedback efficiency, the thermal energy feedback
efficiency is $10^{4}$ times smaller. Thus the coupling between the radiation
and the thermal energy of the gas, on scales between $10^{-2}$ and a
few parsecs, is relatively inefficient.

The efficiency of the AGN energy deposition to the surrounding gas
$\epsilon_{\mathrm{f}}$ is usually treated as a free parameter and is somewhat
constrained by observations, e.g., by fits to
the Magorrian relation and the $M_{\mathrm{BH}}-\sigma$ relations.
SDH05, \citet{Robertson:2006}, \citet{Sijacki:2007}, \citet{DiMatteo:2008}
and \citet{Johansson:2009} find that their numerical simulations
with $\epsilon_{\mathrm{f}}=0.05$ match observations. BS09, who adopted
a slightly different BH accretion model (\S~\ref{sub:Mdot-SPH}),
find a larger value $\epsilon_{\mathrm{f}}=0.15$ produces
a good match with observations. 
In their models, the energy is assumed to be released only in the form of
thermal energy; hence, their $\epsilon_{\mathrm{f}}$ is equivalent to
our $\epsilon_{\mathrm{th}}$. Our thermal energy
feedback efficiency $\epsilon_{\mathrm{th}}$ is about $5\times10^{3}$
times smaller than their $\epsilon_{\mathrm{f}}$.  Even when we
compare it with our total energy feedback efficiency
$\epsilon_{\mathrm{t}}$, their $\epsilon_{\mathrm{f}}$ is still about
$5\times10^{2}$ times larger.

The relatively small value of the kinetic energy feedback efficiency
$\epsilon_{\mathrm{k}}$ found here ($\sim10^{-4}$) 
is consistent with those found in \citet{Chelouche:2008} who analyzed
the X-ray spectra (\emph{Chandra/HETG}) of five AGNs (type-I) with
$M_{\mathrm{BH}}\sim10^{7}\,\Msun$. He found  $\epsilon_{\mathrm{k}}$
ranges between $10^{-6}$ and $10^{-3}$. 
Our models also agree with the relatively low energy and mass feedback
efficiencies found by \citet{Krongold:2007} and  \citet{Stoll:2009}
who studied of X-ray and UV absorbing outflows in Seyfert galaxies.
On the other hand, the AGN evolution and the BH growth synthesis model
of \citet{Merloni:2008}, combined with the observed star formation
rate history of the universe, suggests that
$\epsilon_{\mathrm{k}}=\left(3-5\right)\times10^{-3}$, which is about
an order of magnitude larger than the value found in our models.

While SDH05 and BS09 assumed that the energy feedback efficiency is
constant (independent of accretion luminosity), recent studies by
\citet{Ciotti:2009} and \citet{Shin:2009}, who performed
one-dimensional hydrodynamical simulations of co-evolution of SMBH
and elliptical galaxy,  include the dependency of
the AGN feedback efficiency (in a form of kinetic energy) on the
accretion luminosity or more precisely on the Eddington ratio
($\Gamma$) of the system.  Their parametrization of
$\epsilon_{\mathrm{k}}$ with $\Gamma$ somewhat resembles our
result (Fig.~\ref{fig:eff_energy}), in a sense that
$\epsilon_{\mathrm{k}}$ monotonously increases for $\Gamma \simless
1$.  \citet{Ciotti:2009} find that 
models with varying $\epsilon_{\mathrm{k}}$ are in general more
consistent with observations than those with a constant
$\epsilon_{\mathrm{k}}$.  Interestingly, $\epsilon_{\mathrm{k}}$
values found by \citet{Shin:2009} are only about 5 to 10 times
larger\footnote{The feedback efficiency defined by
  \citet{Ciotti:2009} and \citet{Shin:2009} is equivalent to our
  $\epsilon_{\mathrm{r}} \times \epsilon_{\mathrm{k}}$ where
  $\epsilon_{\mathrm{r}}\approx 0.1$ 
  used in our model.} than our values. This is a much better agreement 
than that with the 
models of SDH05 and BS09, as mentioned above. 
In summary, \citet{Ciotti:2009} and \citet{Shin:2009} concluded that the
models need both radiation and mechanical feedback mechanisms included
at the same time to account for observations such as the ratio of the
central BH mass to the stellar mass ratio ($M_{\mathrm{BH}}/M_{*}$), and the
X-ray luminosity of hot diffused gas.

\subsection{\textcolor{black}{Causes of Low Feedback Efficiencies}}
\label{subsec:cause-low-eff}

\textcolor{black}{
  Relatively low feedback efficiencies found in our
  models (\S~\ref{sub:efficiency-HD}) are mainly caused by the combinations of the
  following: (1)~the axi-symmetric nature of the assumed underlying
  accretion luminosity (the disk radiation), (2)~the axi-symmetric nature
  of the inflow-outflow geometry, and (3)~the relatively low optical
  depth in the outflow. 
  When the Eddington ratio ($\Gamma$) of the system is small, the
  inflow is nearly spherical. However, when the accretion rate is
  increased and $\Gamma$ becomes close to 1, the flow becomes very
  aspherical (cf.~Fig.~\ref{fig:samples}). The inflow occurs only
  near the equator and the outflow occurs in the polar direction.
  Despite a rather large value of $\Gamma$, the accretion can proceed
  in the equatorial region. This is because of the angular
  distribution of the radiation: in our model, the most of the radiation is
  emitted by a flat disk. 
  Consequently, 
  the disk radiation has $\cos{\theta}$ dependency due to the
  geometrical foreshortening, i.e., the flux peaks in
  the polar direction and decreases toward the
  equator. The Eddington ratio, $\Gamma$, is a global property of a system.
  The ratio ($Q$) between the radiation force ($f_{\mathrm{r}}$) and the gravitational
  force ($f_{\mathrm{g}}$), i.e. $Q = f_\mathrm{r}/f_\mathrm{g}$ does not depend on the polar angle
  ($\theta$) for a spherically symmetric case; however, it does depend on
  $\theta$ for a case in which radiation is emitted from flatten
  source.  In the spherical case, when
  $\Gamma>1$, $Q$ is always greater than 1 and an outflow forms in
  all directions ($\theta$). N.B.~for the 
  argument here, we consider the radiation force due to electron
  scattering only, but not including the force due to line
  processes (which is included in our simulations).  For the disk
  radiation case, $Q$ depends on $\theta$,  and $Q$ can be less than 1
  (inflow can occur) near the equator even if the global parameter
  $\Gamma$ is greater than 1.  In our simulations, $Q>1$ in the polar direction (forming
  outflow) and $Q<1$ in the equatorial direction (forming inflow) for
  all the models presented here, except for Model~35 which does not
  form an outflow. 
}
   
\textcolor{black}{
  Once the inflow-outflow geometry becomes very aspherical (i.e.,
  equatorial) as in the middle and right panels of
  Figure~{\ref{fig:samples}}, the coupling between the matter and
  radiation in the polar funnel is reduced. Most of the (high
  density) gas accretes near the equator where the radiation is weakest,
  whereas most of the radiation escapes (without interacting with gas)
  in the polar directions where there is very little accreting
  matter.  The electron optical depth in the outflow in our models
  are rather small ($\tau_{\mathrm{es}} < 1$) even in the highest
  mass-accretion rate model.  This mismatch between the preferred
  direction of the disk radiation and the direction of accretion is a
  main cause of the relatively low feedback efficiencies found in our
  simulations. This occurs in a equatorial system, but not in a
  spherically symmetric system.
}

\textcolor{black}{
    In our models, the outflow is formed from initially inflowing gas.
    In other words, some fraction of the inflowing gas is turned
    around by the radiation force and becomes an outflow.  The
    turn-around points or wind launching points can occur at a large
    distance from the center, i.e., near the outer boundary
    (cf.~Fig.~\ref{fig:samples}).  When the wind is launched from a
    larger radius, the gas does not have time or space to accelerate
    to a terminal velocity before escaping from the outer boundary.
    This may result in slightly smaller momentum and energy feedback
    efficiencies than those obtained from larger scale simulations.
    The efficiencies reported here is strictly on $\sim 0.01$ to $\sim
    10$~pc scale.  Finally, in the simulations presented by KP09, only
    gas dynamics and its microphysics are considered.
    Including the effect of dust would be very important in a larger
    scale simulation ($>10$~pc).  If dust dynamics is included in the
    model and the size of the computational domain is increased, the
    energy and momentum feedback efficiencies could be larger than the
    values reported in this paper.
}

\subsection{\textcolor{black}{Assumption on Mass Accretion Rate}}
\label{subsec:Mdot_a_explain}
\textcolor{black}{
  To compute the total luminosity of the system (Eq.~\ref{eq:L-acc}),
  we need to have the information of the total mass accretion rate
  ($\dot{M_{\mathrm{a}}}$) onto the BH. Since the inner radius
  ($r_\mathrm{i} \sim 0.01$~pc) of the computational domain in the
  simulations in KP09 is much larger than the inner radius of the
  accretion disk (which is on the order of the Schwarzschild radius), we
  assumed that $\dot{M_{\mathrm{a}}}$ is equal to the mass-inflow rate
  at the inner radius of the computational domain
  ($\dot{M_{\mathrm{in}}}\left [ r_\mathrm{i} \right ]$).  In other
  words, we assumed that all the gas which crosses the inner boundary
  will eventually reaches the central black hole.  We also assumed that
  no material can enter the computational domain from the inner
  boundary, whereas the energy and momentum in the form of the
  radiation can.  These assumptions had to be made 
  for numerical reasons. We simply do not know how much gas would
  enter from the inner boundary without explicitly modeling the
  accretion disk and its wind, consistently with the larger scale
  flow.  Unfortunately, the current computer speeds and resources
  would not allow us to include such small scales and the large scale
  ($\sim 10$~pc) flows at the same time in this type of numerical
  simulations.  However, in reality, some gas would enter the
  computational domain from the inner boundary, perhaps due to a disk
  wind (e.g.~\citealt{Proga:2000}) or a jet produced in the immediate
  vicinity of the BH. We simply do not know how much and in what form
  of material enter from the inner boundary; hence, we made the
  simplest assumption: no gas enters into the computational domain
  from the inner boundary.
}

\textcolor{black}{
  We recognize that the feedback from the innermost part of the
  accretion flow may be very important, and may influence the result
  of larger scale simulations such as the ones presented in this
  paper.  Once again, we remind readers that the feedback efficiencies
  reported here are those of large scales (specifically $\sim0.01$ to
  $\sim10$~pc scales), but not in the AGN as a whole (including the
  accretion disk wind/jet).  On the small scale, different forms of
  feedback mechanism (e.g. disk wind and jet) may be important, but
  this is beyond the scope of the current investigation.
}

\section{Conclusions}
\label{sec:Conclusions}

We have presented and analyzed the AGN feedback efficiencies of
energy, momentum and mass based on our axisymmetric and time-dependent
hydrodynamical simulations (see Tab.~\ref{tab:Model-Summary})
presented in \citet{Kurosawa:2009b}. The simulations
capture the
radiation-driven outflows formed from a slowly rotating
(sub-Keplerian) infalling gas under the influence of the gravity of the
central SMBH.  The accretion-luminosity and the outer boundary
temperature are self-consistently determined in these
models. The radial range of the simulations spans from $\sim
10^{-2}$ to $\sim 10$~pc.

The dependency of the mass-accretion rate on the density of the
surrounding gas (or the outer boundary density $\rho_{o}$) has been
examined (Fig.~\ref{fig:mdot}).  The result is compared with the Bondi
mass-accretion rate (Eq.~[\ref{eq:mdot_bondi}]), and with those adopted in
the cosmological simulations of SDH05 and BS09 (see also
Fig.~\ref{fig:mdot-models}).  The density dependency of the
mass-accretion rate in our models is somewhat similar to that of the
Bondi accretion model.  For the density range of $10^{-21} \simless
\rho_{o} \simless 10^{-18}\,\mathrm{g cm^{-3}}$ (or correspondingly for $0.2 \simless
\Gamma \simless 5$), the differences
between the two models are within a factor of 10.  At $\Gamma \approx
1$ ($\rho_{o} \approx 4 \times 10^{-20}\,\mathrm{g cm^{-3}}$), the
accretion rate of our model and that of the Bondi accretion model
agree with each other.  An important difference between the two models
is the steepness of the dependency on the density. The power-law fit
($\dot{M}_{\mathrm{a}} \propto \rho_{\mathrm{o}}^{q}$) of our models
results yields $q\approx 0.5$ instead of $q=1.0$ which is predicted by
the Bondi accretion model. This difference is due to outflows in our model.  
We note that in this density range, the mass-accretion rates 
adopted by SDH05 and BS09 have no density dependency because their
accretion rates are limited by the Eddington rate. The accretion rates
of SDH05 are artificially boosted up ($\alpha$ factor in
Eq.~[\ref{eq:mdot_bondi_sph}]) by a factor of 100 in comparison with
the Bondi mass-accretion rates. Consequently, their rates reach the
Eddington limit at much smaller $\rho_{o}$ than in our simulations
(see Fig.~\ref{fig:mdot-models}).

We find the energy feedback efficiency of our models depends on the
accretion luminosity of the system (Fig.~\ref{fig:eff_energy}). 
For $\Gamma \simless 1$, the dependency is similar
to the parametrization of the feedback efficiency adopted by
\citet{Ciotti:2009}.  Both kinetic and thermal energy efficiencies
($\epsilon_{\mathrm{k}}$ and $\epsilon_{\mathrm{th}}$) peak at around
$\Gamma=1$. The maximum efficiency values are
$\epsilon_{\mathrm{k}}\approx 10^{-4}$ and
$\epsilon_{\mathrm{th}}\approx 10^{-5}$ respectively (see
Table~\ref{tab:efficiency}). For systems with relatively low Eddington
ratio ($\Gamma$$\simless0.4$), the thermal feedback efficiency is higher than the
kinetic feedback efficiency ($\epsilon_{\mathrm{th}}>\epsilon_{\mathrm{k}}$).  On
the other hand, for systems with relatively high Eddington ratio
($\Gamma$$\simgreat0.6$), the kinetic feedback dominates the thermal
feedback by a factor of $\sim10$ to $\sim100$.

The dependency of the momentum feedback efficiency
$\epsilon_{\mathrm{p}}$ on $\Gamma$ is similar to that of the energy
feedback efficiency (Fig.~\ref{fig:eff_momentum}).  
The maximum efficiency is $\epsilon_{\mathrm{p}}\approx 10^{-2}$ which
is about $\sim 100$ times larger than that of the total energy feedback
efficiency.  The dependency of the mass feedback efficiency
$\epsilon_{\mathrm{m}}$ on $\Gamma$ is also similar to that of
$\epsilon_{\mathrm{p}}$ (Fig.~\ref{fig:eff_mass}). 
The maximum value of $\epsilon_{\mathrm{m}}$ found is $\sim0.4$ at
around $\Gamma=1$, i.e., about $40\%$ of the total mass that moves
inward at large radii,
 does not reach the inner boundary of our computational domain, but rather
is turned into outflows. 

Compared to the energy (thermal only) feedback efficiencies
($\epsilon_{\mathrm{f}}=0.05$) required in the recent
cosmological and galaxy mergers simulations (e.g., SDH05;
\citealt{Robertson:2006}, \citealt{Sijacki:2007}, \citealt{DiMatteo:2008}
and \citealt{Johansson:2009}), our thermal energy feedback efficiency
$\epsilon_{\mathrm{th}}$ at the peak value is about $5\times10^{3}$
times smaller than their $\epsilon_{\mathrm{f}}$.  Our total and
kinetic energy efficiencies are about $5\times10^{2}$ times smaller
than their values. These large discrepancies would suggest a few things.
For example, our models are missing important elements. 
In particular, we do not include effects of dust which could make 
the outflows stronger. In addition,
we focus here on axisymmetric models which could
differ from fully three-dimensional (3-D) models. Our preliminary
3-D simulations show that in 3-D, the wind kinetic energy
is smaller than in 2-D while the opposite is true for 
the thermal energy (\citealt{Kurosawa:2009}). Although these changes are small (less than a
factor of 2) in one of the cases studied by \citet{Kurosawa:2009}, 
they could be more significant in other cases (i.e., for the luminosity
higher and lower than $\Gamma=0.6$ assumed by \citealt{Kurosawa:2009}).

It is also possible that the AGN feedback may not be as effective 
as one might have had expected. Instead, other forms of feedback 
may be more significant than the AGN feedback via radiation 
on scales between $10^{-2}$ and a few parsecs. They include,
the supernova feedback, the
radiative feedback from star formation, the strong stellar wind
from massive stars, and strong accretion disk winds or jets from AGN.
The last two forms will introduce magnetic fields
which may carry outward fluxes of energy and momentum.
It is also possible that the AGN feedback efficiencies are indeed low
and the AGNs take a long time to influence their environment. We note
that in our models the AGNs do not shut off the mass supply completely even at 
very high luminosities.
This indicates that the AGNs could operate on a very long time scale
over which their impact on the environment can accumulate, and eventually become
significant. 

Finally, we conclude by noting that AGN feedback has two distinct
modes: (1)~radiation-driven (quasar) mode and (2)~magnetic driven (radio
jet) mode.  In the current cosmological simulations, it is
particularly difficult to deal with the latter, as it requires a full
magnetohydrodynamical (MHD) treatment, and we do not have
an adequate resolution and more importantly the full understanding of
the jet mechanism. Nevertheless, \citet{Sijacki:2007} extended 
the original implementation of the quasar mode feedback in SDH05 by
injecting energy at random positions within a sphere centered around a
BH to emulate the hot gas bubbles created by AGN jets. The radio mode could
be very important in certain situations, such as the cooling flows in
clusters of galaxies where channeling the energy within a very narrow
jet helps to transport the energy outside a galaxy. However, this
occurs in the low accretion rate regime, and it is subdominant in
terms of the total BH mass growth. On the other hand, our
simulations do not include the MHD treatment of the collimated jet;
hence, the focus of this paper is on the physical mechanism of the
quasar mode feedback.  Since the quasar mode is the dominant process
for the total BH mass growth, our results of low feedback efficiencies
would have significant implications on the mass growth of SMBHs in the
early universe. 
In the near future, we plan to further investigate the implications of
our results by taking the boundary conditions of our simulations from
full cosmological hydrodynamic simulations, and thereby making more
direct connections with the physical conditions in a cosmological
context.

\acknowledgements{}
Authors thank the anonymous referee for constructive comments and
suggestions for improving the clarity of the manuscript. We thank
J. M. Stone for suggesting to carry out this study. We also thank
J. Ostriker, J.-H. Choi, and L. Ciotti for useful discussions.  This
work was supported by NASA through grant HST-AR-11276 from the Space
Telescope Science Institute, which is operated by the Association of
Universities for Research in Astronomy, Inc., under NASA contract
NAS5-26555. We acknowledge support from NSF (grant AST-0807491) and
the National Aeronautics and Space Administration under
grant/Cooperative Agreement No.~NNX08AE57A issued by the Nevada NASA
EPSCoR program.



\end{document}